\documentclass[aps,prl,twocolumn,showpacs]{revtex4}

\usepackage{graphicx}

\def\vR{{\bf R}}

\def\vk{{\bf k}}

\begin{document}

\title{Multiband tight-binding model of local magnetism in Ga$_{1-x}$Mn$_x$As}

\author{Jian-Ming Tang and Michael E. Flatt\'e}

\affiliation{Department of Physics and Astronomy, University of Iowa,
  Iowa City, IA 52242-1479}


\begin{abstract}

  We present the spin and orbitally resolved local density of states
  (LDOS) for a single Mn impurity and for two nearby Mn impurities in
  GaAs. The GaAs host is described by a $sp^3$ tight-binding
  Hamiltonian, and the Mn impurity is described by a local $p$-$d$
  hybridization and on-site potential.  Local spin-polarized
  resonances within the valence bands significantly enhance the LDOS
  near the band edge. For two nearby parallel Mn moments the acceptor
  states hybridize and split in energy. Thus scanning tunneling
  spectroscopy can directly measure the Mn-Mn interaction as a
  function of distance.

\end{abstract}


\maketitle

The successful growth of ferromagnetic diluted magnetic semiconductors
(DMSs) based on III-V
compounds~\cite{Munekata1989,Ohno1992,Medvedkin2000,Reed2001,Sonoda2002}
has attracted much attention due to the nearby metal-insulator
transition and also for their potential application to nano-scale
nonvolatile storage devices and to quantum
computation~\cite{Wolf2001,Awschalom2002}. In order to turn a
nonmagnetic semiconductor into a magnet, a sizable amount ($\sim 1\%$)
of magnetic dopants are introduced.  Experimental
studies~\cite{Ohno1992,Matsukura1998,Szczytko1999a,Beschoten1999,Ohno2000,Nagai2001}
have shown strong correlations between the ferromagnetism and the hole
carriers that are also contributed by the Mn doping in the III-V DMSs.
The emerging picture of this ferromagnetism is that the Mn magnetic
moments are localized, and the ferromagnetic coupling is mediated
through the delocalized hole carriers~\cite{Dietl2000,Awschalom2002}.
A deeper understanding of this ferromagnetic coupling has been
hampered by insufficient knowledge of the valence band structure. The
assumption that the holes reside in the unperturbed valence bands of
the host semiconductors~\cite{Ohno1992,Dietl1997,Matsukura1998} has
recently been questioned by both theoretical and experimental
studies~\cite{Akai1998,Beschoten1999,Nagai2001,Okabayashi2001}.
Effective mass theories provide good spectral resolution at the band
edge, but cannot describe distortions on distances of one or two
lattice constants for they are only accurate for momenta close to the
$\Gamma$ point.  Density functional
calculations~\cite{Sanvito2001,Schilfgaarde2001}, however, describe
local properties well, but state of the art supercell calculations
have not had sufficient spectral resolution to resolve the shallow
bound states in the gap and the sharp resonances in the valence bands.

\begin{figure}
\centerline{ \includegraphics[width=\columnwidth]{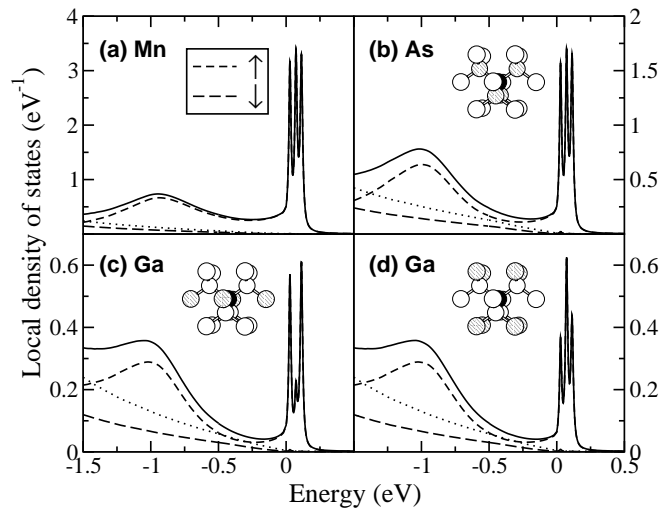} }
\caption{ LDOS spectra for the charged (ionized) Mn dopant in
  Ga$_{1-x}$Mn$_x$As (corresponding to bulk {\em n}-doped) at (a) the
  Mn site, at (b) one of the nearest-neighbor As sites, and at (c and
  d) one of the second-nearest-neighbor Ga sites. The shaded balls in
  each panel show the corresponding atomic sites, and the black ball
  shows the Mn site.  The Mn core spin is aligned with the $z$-axis of
  the lattice, and the $z$-axis is aligned vertically. In this
  particular case, the spectra at four nearest-neighbor sites are
  identical, and have two different functional forms at the
  second-nearest-neighbor sites. The total LDOS is shown by the solid
  lines. The dashed lines show the projections onto the two spin
  polarizations relative to the Mn core spin. The dotted lines show
  the LDOS of the bulk GaAs. The zero of energy is the valence maximum
  of bulk GaAs.  The valence band LDOS of bulk GaAs is more
  concentrated at the As sites than at the Ga sites, because the LDOS
  is the density of states modulated by the probability density of
  electrons. For the neutral Mn dopant (corresponding to bulk {\em
    p}-doped Ga$_{1-x}$Mn$_x$As) only the highest energy peak of the
  three localized peaks apparent above in each panel would be
  visible.}
\label{fig:spectra}
\end{figure}

In this Letter we study the local density of states (LDOS) of very
dilute concentrations of Mn in GaAs as a prototype III-V DMS, and show
that the valence bands are strongly altered by the Mn dopants, and in
return influence the interaction between the Mn magnetic moments.  In
order to obtain both sufficient spectral resolution and proper
dispersion relations throughout the full Brillouin zone, a multiband
tight-binding approach incorporating spin-orbit interaction is
employed. Our results show that the hybridization between the Mn
$3d$-orbitals and the GaAs valence bands leads to spin-polarized
resonances {\it within the valence bands} and to delocalized
ferromagnetic interaction.  The LDOS near the valence band edge is
significantly enhanced by these resonances. Each Mn dopant can enhance
the LDOS by as much as a factor of $2$ up to the second-nearest
neighbors (See Fig.~\ref{fig:spectra}), corresponding to a $\sim 7\%$
increase of the average LDOS for $1\%$ Mn concentration.  The strong
LDOS enhancement is consistent with recent angle-resolved
photoemission measurements~\cite{Okabayashi2001}, and qualitatively
explains the increases of the absorption coefficients in the
intraband~\cite{Nagai2001} as well as the interband
absorption spectroscopies~\cite{Szczytko1999a,Beschoten1999}. In
particular, the experimental result that the band-edge absorption
coefficient increases in finite magnetic fields more for one
polarization of light is consistent with our result that the
resonances are both orbitally- and spin-polarized.

Whereas the spin-polarized band-edge resonances will alter the Mn-Mn
interaction, the acceptor states provide a direct means of
measuring that interaction. The acceptor level of Mn splits when two
Mn magnetic moments are in a parallel configuration~\cite{Flatte2000}.
The range of interaction between the Mn dopants can be probed by the
size of the splitting as a function of the separation. We find that the
splittings are as large as tens of meV even when two
Mn dopants are separated by a few lattice constants. This magnetic
interaction is anisotropic with respect to the axis connecting the two
Mn dopants due to spin-orbit interaction, in qualitative agreement
with the continuum model~\cite{Zarand2002}. Both the local enhancement
of the valence band edge and the splitting of the acceptor level could
be probed by scanning tunneling spectroscopy
(STS)~\cite{Feenstra2002}.

As pointed out by \citet{Vogl1985}, the fivefold degenerate Mn
$3d$-orbitals split in a cubic lattice to an $E$-symmetric
($d_{x^2-y^2}$-like) doublet, which is only weakly coupled to the
tetrahedral host, and a $T_2$-symmetric ($d_{xy}$-like) triplet, which
can effectively couple to the neighboring dangling $sp^3$-hybrids.
The antibonding states of the $T_2$-symmetric states and the
$sp^3$-hybrids form the acceptor states within the gap. These
antibonding states are delocalized in space since they overlap
strongly with the host valence bands. In our calculations the
hybridization of the majority Mn $d$-orbitals is treated as an
effective extended spin-dependent potential $U_1$ acting on the four
nearest-neighbor sites of Mn, because only $s$ and $p$ orbitals are
explicitly used in our tight-binding Hamiltonian.  The operator form
of our Mn potential is
\begin{eqnarray}
V & = & U_0\sum_{\ell,s}c^\dagger_{\ell s}(\vR_0)c_{\ell s}(\vR_0) + U_1\sum_{j=1}^4\sum_\ell c^\dagger_{\ell\uparrow}(\vR_j)c_{\ell\uparrow}(\vR_j) \;,\nonumber\\
\end{eqnarray}
where $U_0$ is the on-site orbital energy difference, $c^\dagger_{\ell
  s}(\vR)$ ($c_{\ell s}(\vR)$) is the creation (annihilation) operator
of a spin-$s$ electron in the $\ell$ orbital at site $\vR$. The Mn
dopant is located at $\vR_0$, and the four nearest-neighbor sites are
labeled by $\vR_1$-$\vR_4$. We neglect the difference of the potential
matrix elements between the $s$ and $p$ orbitals, because the
$s$-orbitals make a negligible contribution to the LDOS near the
valence band edge.  The quantization axis for the spin is aligned with
the Mn core spin.  The minority Mn d-orbitals are assumed to be much
higher in energy, and the hybridization energy of them with the
spin-down states is neglected.  We find that the effective potential
$U_1$ is of critical importance in inducing the acceptor level at the
experimentally observed energy.  The significance of the hybridization
to the acceptor level energy has also been noted recently by
\citet{Dietl2002}. As for the delocalized nature of the acceptor
state, our results show that only about $10\%$ of the spectral weight
of the acceptor state is concentrated at the Mn site, $20\%$ is
distributed over the four nearest-neighbor sites, and the remaining
$70\%$ is extended to farther sites.  Furthermore, the large
spin-orbit interaction of GaAs splits the gap states into three
different energy levels, and introduces anisotropy to the spatial
structure.

We use the Koster-Slater technique~\cite{Koster1954} to calculate the
Green's function whose imaginary part gives the LDOS. This method has
been proven to give the correct chemical trend of impurity levels in
semiconductors~\cite{Hjalmarson1980}, and we have previously applied
it to successfully predict STM spectra near impurities in
superconductors~\cite{Tang2002}. Starting with the tight-binding
Hamiltonian $\hat H_0(\vk)$ of homogeneous GaAs, one first calculates
the retarded Green's function, $\hat G_0(\vk,\omega)=[\omega-\hat
H_0(\vk)+i\delta]^{-1}$. To obtain a good description of the valence
band structure, we use the $sp^3$ model including spin-orbit
interaction~\cite{Chadi1977}. Then, by Fourier transforming $\hat
G_0(\vk,\omega)$, we construct the homogeneous Green's function in
coordinate space, $\hat G_0(\vR_i,\vR_j,\omega)$, where $\vR_i$ and
$\vR_j$ label the zincblende lattice sites. This step consumes the
majority of computation time. It takes about one day for one link
($\vR_i-\vR_j$) with a spectral range of $2$ eV on a personal
computer.  The number of links used in this study is about 500.  A
constant linewidth of $\delta=10$ meV produces a good spectral
resolution within a reasonable computation time. The final Green's
function is obtained by solving the Dyson's equation,
\begin{eqnarray}
\check G(\omega) & = & \check G_0(\omega) + \left[\check{\bf 1}-\check G_0(\omega)\check V\right]^{-1}\check G_0(\omega) \;,
\end{eqnarray}
where $\check G(\omega)$ is the full matrix representation using all atomic
orbitals at all lattice sites. The LDOS at each site $\vR_i$ is given
by
\begin{eqnarray}
A(\vR_i,\omega) & = & -\frac{1}{\pi}{\rm Im}\left[{\rm tr}_\alpha\,\hat G(\vR_i,\vR_i,\omega)\right] \;,
\end{eqnarray}
where the ${\rm tr}_\alpha$ is taken with respect to the orbitals of the
$\alpha$ atom, depending on which type of atom is actually located at the
site $\vR_i$.

To determine the values of $U_0$ and $U_1$, we first consider the
possibility that the potential is nonzero only at the Mn site
($U_1=0$), and the size of the on-site potential is assumed to be the
same for all orbitals. When the Mn atom replaces a Ga atom, the
strength of this on-site potential is estimated to be $U_0\sim 1$ eV
based on the energy difference of the ionization energies of Ga
($4s^24p$) and of Mn ($3d^54s4p$), which are about $6$ and $5$ eV
respectively. Not only is this on-site potential too weak to bind any
acceptor level in the gap, it is not possible (even with an
unrealistically large on-site potential, $U_0\gg 10$ eV) to bind a
hole more than $60$ meV above the valence band edge. The Mn dopant
introduces an acceptor level at $113$ meV above the valence band edge,
therefore, the effective potential must be extended at least to the
four nearest-neighbor As sites. If we fix $U_0=1$ eV, and tune $U_1$
to give the correct acceptor level energy, the required
nearest-neighbor potential is about $3.59$ eV.  This nearest-neighbor
potential is too large to be accounted for by the tail of the screened
Coulomb potential.  The screened Coulomb potential at the
nearest-neighbor sites of a singly charged Mn center is only about
$0.5$ eV, or even smaller if the system is doped with carriers, as in
our case.  Therefore, the large effective potential at the
nearest-neighbor sites must result from the hybridization of the Mn
$3d$-orbitals and the host $sp^3$-hybrids. The exchange interaction
effectively pushes the valence-band states with the same spin
polarization as the Mn core spin into the gap. The states with the
opposite spin polarization essentially remain unperturbed.

\begin{figure}
  \centerline{
    \includegraphics[width=\columnwidth]{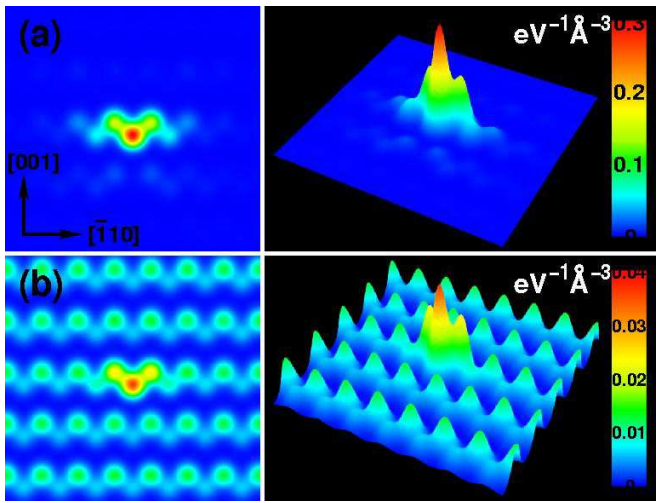}
  }
\caption{ (color) The spatial structure of
  (a) the Mn acceptor level at $113$ meV and of (b) the valence band
  LDOS at $-500$ meV. The LDOS peaks are centered at the Mn atom, and
  are extended mainly on the $(001)$ plane.  In (b), far away from the
  Mn site, the sites with higher LDOS are occupied by the As atom, and
  with lower LDOS are occupied by the Ga atom. We have assumed that
  the squared modulus of the Wannier functions to be a Gaussian with a
  width of half the distance between neighboring Ga and As atoms. }
\label{fig:spatial}
\end{figure}

\begin{figure}
\includegraphics[width=0.9\columnwidth]{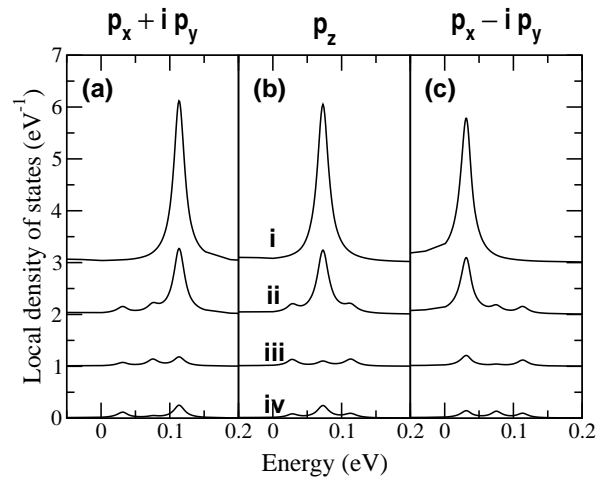}
\caption{ The orbitally resolved LDOS for (a) $p_x+ip_y$, (b) $p_z$
  and (c) $p_x-ip_y$ orbitals. The Mn core spin is aligned with the
  $z$-axis. Each panel contains four LDOS curves.  From top to bottom,
  the curves show the LDOS at (i) the Mn, (ii) the nearest-neighbor
  (As) sites, and (iii and iv) the second-nearest-neighbor (Ga)
  sites. Cases (iii) and (iv) correspond to the configurations (c) and
  (d) in Fig.~\ref{fig:spectra} respectively. The LDOS spectra are
  shifted by multiples of one unit for easier visualization.  }
\label{fig:orbitals}
\end{figure}

Figure~\ref{fig:spectra} shows the LDOS spectra at the Mn site and at
neighboring sites. In this particular case we have chosen the Mn core
spin to align with one of the (100) crystal axes (which is the bulk
easy axis~\cite{Tang2003}). As a result, the spectra at the four
nearest-neighbor As sites are identical due to the residual symmetry
operations of $C_2$ and $S_4$ about the crystal axes. The twelve
second-nearest-neighbor Ga sites are divided into two classes with
four and eight sites each. The acceptor states are almost fully
spin-polarized (parallel to the Mn core spin) and split into three
energy levels due to the spin-orbit interaction. For bulk {\em
n}-doped Ga$_{1-x}$Mn$_x$As the Fermi level lies above the upper
acceptor state, so the impurity is ionized and all three levels would
be visible in a tunneling experiment.  If the Fermi level lies below
the upper acceptor state, such as in bulk {\em p}-doped
Ga$_{1-x}$Mn$_x$As, then due to the electron-electron interaction only
one localized state will be visible above the Fermi level and none
below. Due to band-bending at the surface it is often possible to see
both the ionized and neutral dopants in the same sample (as in
Ref.~\onlinecite{Feenstra2002}). As the Mn core spin rotates, the orbital
character of the acceptor states changes, but the three energy levels
remain the same within the $sp^3$ model. There is some experimental
evidence for the existence of the two additional acceptor energy
levels~\cite{Lee1964}.  The significant enhancement of the spin-up
LDOS continues into the valence band to $1.5$ eV below the band edge.
The spin-polarized resonances emerge at an energy close to the
split-off band top ($\sim 350$ meV below the band edge). On the other
hand, the spin-down band is weakly perturbed and non-resonant.
Figure~\ref{fig:spectra} (c) and (d) also show that the large
enhancement of the valence band LDOS extends to the second-nearest
neighbors. From the spectral weight of the acceptor state, $\sim 5\%$
at each As site, we estimate the $p$-$d$ exchange interaction $\beta
N_0$ to be $\frac{2}{5}U_1\times 20\% \approx -0.3$ eV, within a
factor of $3$ of that obtained from transport
measurements (see Chapter 1 in Ref.~\onlinecite{Awschalom2002}). 
The discrepancy may come from the
enhanced LDOS near the Mn spin, which is not included in the transport
modeling.

The spatial structure of the LDOS is further illustrated in
Fig.~\ref{fig:spatial}. For the Mn potential used here, the upper
acceptor state is peaked at the Mn site. The ratio between the peak
heights at the Mn and at the nearest-neighbor As sites is related to
the strength of $U_0$ and $U_1$. Generally, increasing the potential
strength at one site reduces the spectral weight at the same site.
The degree of acceptor-state localization apparent in
Fig.~\ref{fig:spatial} is determined largely by 
the effective band masses of the GaAs host, which are dependent on the 
tight-binding parameters of Ref.~\onlinecite{Chadi1977}.
If the spin is aligned with the $z$ axis, the uppermost acceptor state is
spatially extended in the $x$-$y$ plane, because the orbital
character, shown in Fig.~\ref{fig:orbitals}, is $p_x+ip_y$.

\begin{figure}
\centerline{\includegraphics[width=\columnwidth]{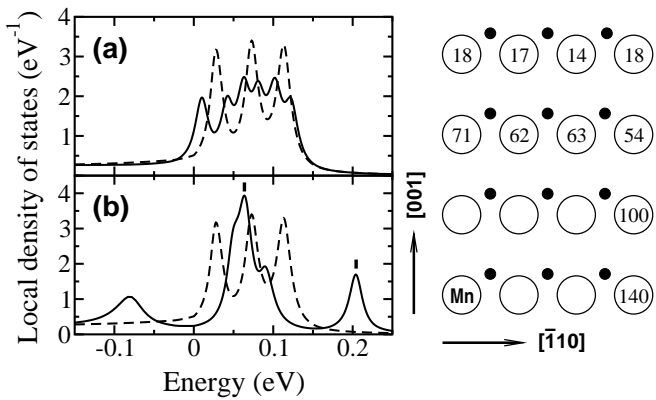}}
\caption{Splitting of the acceptor levels.
  The left panel shows the spectra at the Mn site with two Mn dopants
  separated by (a) $(0,0,3)$ and (b) $(-3/2,3/2,0)$. The dashed line
  shows the spectrum when there is only one Mn dopant. In (b) the two
  split peaks of the $113$ meV level are indicated by the two
  heavy-line segments. The right panel shows the amount of the
  splitting of the $113$ meV level. One of the Mn atoms is located at
  the $(0,0,0)$ site (bottom-left corner), and the other is at one of
  the circles.  The number in the circle shows the splitting energy in meV.
  Both Mn core spins are aligned with the $z$-axis. The black dots
  show the As sites. }
\label{fig:splitting}
\end{figure}

We also calculate the LDOS for two nearby Mn dopants by inverting the
Dyson equation for a two-impurity potential.  When the two Mn core
spins are parallel, the acceptor states centered at the two Mn sites
interfere, and further split into ``bonding'' and ``antibonding''
states. When the two Mn core spins are antiparallel, one of the $U_1$
terms acts on spin-down As-orbitals, and the potential for the spin-up
(down) states at one Mn site is the same as for the spin-down (up)
states at the other Mn site. Therefore, the acceptor states remain in
three degenerate levels.  The splitting of the acceptor level for two
parallel Mn core spins can be used as a probe for the range of
interaction between Mn dopants.  Figure~\ref{fig:splitting} shows the
long-ranged nature of the interaction. It also shows the anisotropy of
the interaction resulting from the spin-orbit interaction. Because the
acceptor state has an orbital character more extended in the plane
perpendicular to the Mn core spin, the splitting of the acceptor level
is the largest when the Mn core spins are oriented perpendicular to
the axis that joins them.

In summary we have presented calculations of the LDOS near Mn in
GaAs. The $p$-$d$ hybridization induces spin-polarized resonances that
enhance the LDOS near the valence band edge. The spin-orbit
interaction splits the acceptor level, and introduces anisotropy to
the Mn-Mn interaction in addition to the crystal anisotropy. The range
of the Mn-Mn interaction is shown to extend through several lattice
constants. For quantitative comparisons with STS measurements,
significant surface effects are expected for a Mn atom at the top
layer because the Mn atom has one less As neighbor, and the
hybridization changes. However, it has been reported that dopant atoms
as deep as in the fifth layer are seen by STS~\cite{Feenstra2002}, and
the spectra for these cases should closely resemble the bulk.

This work was supported under the ARO MURI DAAD19-01-1-0541.

\end{document}